\documentstyle[psfig]{mn}

\newif\ifAMStwofonts
\def\simgt{\hbox{\rlap{\raise 0.425ex\hbox{$>$}}\lower 0.65ex\hbox{$\sim$}}}
\def\simlt{\hbox{\rlap{\raise 0.425ex\hbox{$<$}}\lower 0.65ex\hbox{$\sim$}}}
\def\arcsec{^{\prime\prime}}
\def\arcmin{^\prime}
\def\degree{^\circ}

\input{psfig}

\ifoldfss
  \ifCUPmtlplainloaded \else
    \NewTextAlphabet{textbfit} {cmbxti10} {}
    \NewTextAlphabet{textbfss} {cmssbx10} {}
    \NewMathAlphabet{mathbfit} {cmbxti10} {} % for math mode
    \NewMathAlphabet{mathbfss} {cmssbx10} {} %  "   "    "
  \fi
  \ifAMStwofonts
    \ifCUPmtlplainloaded \else
      \NewSymbolFont{upmath} {eurm10}
      \NewSymbolFont{AMSa} {msam10}
      \NewMathSymbol{\upi}     {0}{upmath}{19}
      \NewMathSymbol{\umu}     {0}{upmath}{16}
      \NewMathSymbol{\upartial}{0}{upmath}{40}
      \NewMathSymbol{\leqslant}{3}{AMSa}{36}
      \NewMathSymbol{\geqslant}{3}{AMSa}{3E}

      \let\leq=\leqslant 
      \let\geq=\geqslant 
    \fi
  \fi
\fi % End of OFSS

\ifnfssone
  \newmathalphabet{\mathit}
  \addtoversion{normal}{\mathit}{cmr}{m}{it}
  \addtoversion{bold}{\mathit}{cmr}{bx}{it}
  \newmathalphabet{\mathbfit} % math mode version of \textbfit{..}
  \addtoversion{normal}{\mathbfit}{cmr}{bx}{it}
  \addtoversion{bold}{\mathbfit}{cmr}{bx}{it}
  \newmathalphabet{\mathbfss} % math mode version of \textbfss{..}
  \addtoversion{normal}{\mathbfss}{cmss}{bx}{n}
  \addtoversion{bold}{\mathbfss}{cmss}{bx}{n}
  \ifAMStwofonts
    \ifCUPmtlplainloaded \else
      \UseAMStwoboldmath
      \makeatletter
      \new@mathgroup\upmath@group
      \define@mathgroup\mv@normal\upmath@group{eur}{m}{n}
      \define@mathgroup\mv@bold\upmath@group{eur}{b}{n}
      \edef\UPM{\hexnumber\upmath@group}
      \new@mathgroup\amsa@group
      \define@mathgroup\mv@normal\amsa@group{msa}{m}{n}
      \define@mathgroup\mv@bold\amsa@group{msa}{m}{n}
      \edef\AMSa{\hexnumber\amsa@group}
      \makeatother
      \mathchardef\upi="0\UPM19
      \mathchardef\umu="0\UPM16
      \mathchardef\upartial="0\UPM40
      \mathchardef\leqslant="3\AMSa36
      \mathchardef\geqslant="3\AMSa3E

      \let\leq=\leqslant 
      \let\geq=\geqslant 
    \fi
  \fi
\fi % End of NFSS release 1

\ifnfsstwo
  \DeclareMathAlphabet{\mathbfit}{OT1}{cmr}{bx}{it}
  \SetMathAlphabet\mathbfit{bold}{OT1}{cmr}{bx}{it}
  \DeclareMathAlphabet{\mathbfss}{OT1}{cmss}{bx}{n}
  \SetMathAlphabet\mathbfss{bold}{OT1}{cmss}{bx}{n}
  \ifAMStwofonts
    \ifCUPmtlplainloaded \else
      \DeclareSymbolFont{UPM}{U}{eur}{m}{n}
      \SetSymbolFont{UPM}{bold}{U}{eur}{b}{n}
      \DeclareSymbolFont{AMSa}{U}{msa}{m}{n}
      \DeclareMathSymbol{\upi}{0}{UPM}{"19}
      \DeclareMathSymbol{\umu}{0}{UPM}{"16}
      \DeclareMathSymbol{\upartial}{0}{UPM}{"40}
      \DeclareMathSymbol{\leqslant}{3}{AMSa}{"36}
      \DeclareMathSymbol{\geqslant}{3}{AMSa}{"3E}

      \let\leq=\leqslant 
      \let\geq=\geqslant 
    \fi
  \fi
\fi % End of NFSS release 2

\ifCUPmtlplainloaded \else
  \ifAMStwofonts \else % If no AMS fonts
    \def\upi{\pi}
    \def\umu{\mu}
    \def\upartial{\partial}
  \fi
\fi

\title[Weak lensing: LBQS and APM]{Angular correlations between LBQS and 
APM: Weak Lensing by the Large Scale Structure}
\author[L. L. R. Williams and M. Irwin]
       {Liliya L. R. Williams$^{1}$\thanks{Email: {\bf \tt llrw@ast.cam.ac.uk}} and Mike Irwin$^{2}$\thanks{Email: {\bf mike@ast.cam.ac.uk }} \\
${^1}$Institute of Astronomy, Madingley Road, Cambridge, CB3 0HA, UK \\
${^2}$Royal Greenwich Observatory, Madingley Road, Cambridge, CB3 0EZ }
\date{MNRAS, in press}
\pubyear{1998}

\begin{document}

\maketitle

\newcommand{\fmmm}[1]{\mbox{$#1$}}
\newcommand{\scnd}{\mbox{\fmmm{''}\hskip-0.3em .}}
\newcommand{\scnp}{\mbox{\fmmm{''}}}

\begin{abstract}
We detect a positive angular correlation between bright, high-redshift 
QSOs and foreground galaxies. The QSOs are taken from the optically 
selected LBQS Catalogue, while the galaxies are from the APM Survey. 
The correlation amplitude is about a few percent on angular scales of 
over a degree. It is a function of QSO redshift and apparent magnitude, 
in a way expected from weak lensing, and inconsistent with QSO-galaxy 
correlations being caused by physical associations, or uneven obscuration 
by Galactic dust. The correlations are ascribed to the weak lensing 
effect of the foreground dark matter, which is traced by the APM galaxies. 
The amplitude of the effect found here is compared to the analytical 
predictions from the literature, and to the predictions of a
phenomenological model, which is based on the observed counts-in-cells 
distribution of APM galaxies. While the latter agree reasonably well 
with the analytical predictions (namely those of Dolag \& Bartelmann 1997, 
and Sanz {\it et al.} 1997), both under-predict the observed correlation 
amplitude on degree angular scales. We consider the possible ways to
reconcile these observations with theory, and discuss the implications
these observations have on some aspects of extragalactic astronomy.

\end{abstract}


\section{Introduction}\label{intro}

It is now widely recognized that the sky-projected density of 
high-redshift objects can be altered due to weak lensing by intervening
mass distribution, on a range of angular scales (Narayan 1989,
Broadhurst, Taylor \& Peacock 1995) The effect arises 
because gravitational lensing distorts the area on the sky in the 
direction of the lens. Behind a lens of a positive mass density, the 
area is stretched out. As a result: (1) individual sources subtend a 
larger area on the sky and therefore look brighter due to flux 
conservation; (2) but their number density decreases. If the slope of 
magnitude number counts is steep then the first effect wins over the 
second and the net number density of background sources in a flux limited 
survey is increased; if the slope is shallow, the number density
is decreased. Thus the amplitude of the effect depends on the amount and 
clumpiness of the lensing matter, the relative redshifts of the lenses 
and sources, and the slope of the source number counts in the appropriate 
redshift range. 

The effect of dilution or enhancement of the projected number density
is most pronounced for sources with magnitude-number counts whose slope 
deviates strongly from $\alpha$=d$logN$/d$m$ of 0.4. In the cases where 
the slope 
is shallower then 0.4, anti-correlations with foreground lenses have been
observed or suspected: Rodrigues-Williams \& Hogan (1994) suggest that the 
anti-correlation of faint UVX objects with clusters (Boyle et al. 1988) 
are due to weak lensing rather than dust. Broadhurst (1994) observe a 
deficit of red faint galaxies ($\alpha\approx$ 0.3) behind the
foreground cluster Abell 1689. 

When the slope is steeper than 0.4, positive correlations are detected.
These observations can be roughly categorized by the angular scale
of correlations. On small scales, $3-30\arcsec$, there are a number 
of observations (see Narayan 1991, and Hewett, Harding \& Webster 1991 
for reviews). Though direct comparison between these is difficult, because 
each sample has its own selection criteria and method of
analysis, the correlations are thought to be reasonably well understood 
in terms of lensing by individual galaxy dark matter halos. 

On $1-15\arcmin$ scales the correlations are due to dark matter
associated with galaxies, or clusters of galaxies, on Mpc scales. 
Such correlations with
radio sources are well documented. $z>0.5$ 1 Jy Catalogue sources have
been cross correlated with almost every available catalogue of `low
redshift' extragalactic objects: ~Lick galaxies (Fugmann 1990,
Bartelmann \& Schneider 1993), ~IRAS galaxies, and ~EMSS (Bartelmann \&
Schneider 1994), ~ROSAT All-Sky Survey (Bartelmann, Schneider, \&
Hasinger 1994), ~Zwicky clusters (Seitz \& Schneider 1995), ~APM
galaxies (Ben\'{\i}tez, N., \& Mart\'{\i}nez-Gonz\'ales 1995, 1997), and
~Abell clusters (Wu \& Han 1995).  In all instances positive
correlations were found, though at varying statistical significance,
from 1.5 to $\simgt$3 $\sigma$. The large amplitude of the effect
remained unexplained until recently. Dolag \& Bartelmann (1997) and
Sanz et al. (1997) reproduced the correlations on $\sim 10\arcmin$
scales by incorporating the non-linear growth of the matter power
spectrum in the Universe. Including these mass fluctuations leads to
an order of magnitude increase in the QSO-galaxy correlations on Mpc
scales. 

On yet larger scales, $20\arcmin-1\degree$, there are three existing
studies with homogeneous complete catalogues of objects. 
Rodrigues-Williams \& Hogan (1994) looked at correlations between LBQS 
QSOs (Hewett et al. 1995) and Zwicky galaxy clusters, and found a 
significant signal with a subset of QSOs at $1.4\leq z\leq 2.2$. Seitz \&
Schneider (1995) extended the study to 1 Jy radio sources, and a range
of QSO subsamples based on redshift and apparent magnitude; sources
at $z\sim 1$ were found to be associated with foreground clusters
at $97.7\%$ significance level. While Rodrigues-Williams \& Hogan argue 
that the correlations could not be the result of patchy
obscuration by the Galactic dust, the study of Seitz \& Schneider,
who used radio selected sources, proved so beyond a doubt. The physical
nature of associations can be definitely ruled out based on the
discrepant redshifts of QSOs and clusters, leaving weak gravitational 
lensing as the only plausible explanation. Ferreras et al. (1997)
detect a strong QSO-galaxy anticorrelation between faint, $z<1.6$, 
optically  selected QSOs in a 5.5 deg$^2$ region close to the North 
Galactic Pole. The authors attribute the effect to the selected biases 
associated with identifying QSOs in crowded areas.

In this paper we extend the previous work by examining the correlations 
between optically selected QSOs and foreground galaxies, on scales 
of up to a degree. In the following sections we measure the QSO-galaxy 
cross-correlation signal (Section~\ref{observations}), compare it to 
theoretical predictions (Section~\ref{analysis}), and discuss a few 
areas of cosmology that our observations will have a considerable impact 
on (Sections~\ref{discussion}, and ~\ref{implications}).

\section{Data}\label{data}

The Large Bright Quasar Survey (LBQS, Hewett et al. 1995) is the largest 
homogeneous catalogue of optically selected QSOs. Candidates were obtained
from machine-scanned direct and objective-prism UK Schmidt Telescope 
plates, based on several selection criteria, including blue colour excess 
and presence of strong emission lines. The final list of QSOs was compiled 
after spectroscopic follow-up of the candidates at the Multiple Mirror
and other telescopes. The LBQS contains over a 1000 QSOs between $m_B=16$ 
and $\approx 
18.5$ in 18 high Galactic latitude fields. The QSO redshift distribution, 
presented in Figure 9 of Hewett et al. (1995), is smooth and contains no 
gaps in the redshift range between 0.2 to 3.5. In the present work we only 
use the eleven equatorial LBQS fields. We do not use four fields in
the direction of the Virgo Cluster, because the faint galaxy counts in the 
these fields can be severely contaminated by the Virgo galaxies. We do 
not use two fields in the direction of the Galactic bulge, because of star
contamination. We also did not use one field near the South Galactic pole.

The APM Catalogue (Irwin et al. 1994) was compiled from the scans 
of Schmidt plates, carried out by the Automatic Plate Measuring facility 
at Cambridge. The Catalogue is a list of objects, detected on red and 
blue plates separately, and classified based on their morphology as 
star-like, extended, noise, or blend. The galaxy data we use is only
approximately magnitude calibrated; two fields can be offset by as much 
as one magnitude. However, we do not attempt to correct for this because 
the difference is not large for our purposes, and our analysis compensates 
for it. 

Each Schmidt plate is $\approx 6\degree$ across, but in order to reduce 
vignetting problems we only use objects located within 2.7$\degree$ of 
the plate centre.

\begin{figure}
\vbox{
\centerline{
\psfig{figure=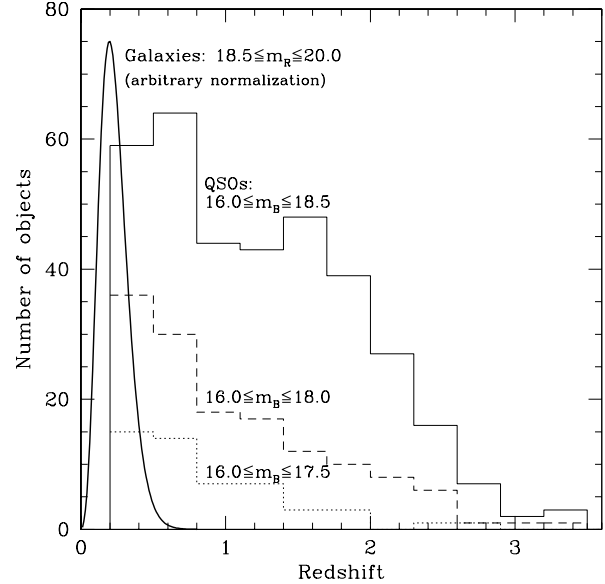,width=3.25in,angle=0}
}
\caption[]{
Redshift distribution of the objects used in the present work. The three
histograms are for QSOs with $m_B\leq$ 17.5, 18.0, and 18.5, all taken 
from eleven LBQS equatorial fields. The solid line is the estimated 
redshift distribution (Maddox et al. 1996) for $18.5\leq m_R\leq 20.0$ 
APM galaxies (arbitrary vertical normalization). 
\label{z}
}
}
\end{figure}

Figure~\ref{z} shows the redshift distribution of LBQS QSOs and the APM 
galaxies. The QSOs (three histograms for $m_B\leq$ 17.5, 18.0, and 18.5)
are those found in 11 equatorial fields used in this work.
The solid curve represents
$18.5\leq m_R\leq 20.0$ galaxies, and is based on the redshift
distribution estimated by Maddox et al. (1996). Because the latter
analysis applies to APM galaxies detected on the blue plates while 
we mostly use red plates, we assumed a uniform colour transformation 
of $B-R=1.5$, which is the average colour of the $18.5\leq m_R\leq 20.0$ 
galaxies detected on both red and blue APM plates. It is also consistent
with the $B-R$ colour derived by Metcalfe et al. (1995). The peak of 
the galaxy redshift distribution is at $z\approx 0.2$, and virtually no 
galaxies lie beyond $z=0.7$.

\section{QSO-galaxy Correlations}\label{observations}

In this Section we ask ourselves if the sky-projected distributions
of the APM galaxies and background LBQS QSOs are correlated.
The amplitude of the signal, if any, is expected to be small, 
therefore it is important to take into account any possible biases
that may affect the signal. In particular, Schmidt plates suffer
from radial sensitivity gradients, mainly caused by vignetting,
which results in radially dependent 
object number density. These gradients are small, and can be different 
for stars vs. galaxies, and for faint vs. bright objects. Our
cross-correlation analysis, which we describe in the next Section,
takes these effects into account.

\subsection{Method---cross-correlation estimator}\label{estimator}

The standard measure of the clustering of objects is the
two-point correlation function. The commonly considered case, the
auto-correlation function of galaxies, has been explored in the
literature in great depth; Peebles (1980), Hewett (1982), Hamilton (1993)
have derived estimators that can tackle various observational
limitations of the data, like finite field size, average density 
variations on large angular scales, plate inhomogeneities, etc. 
Our case, QSO-galaxy cross-correlation, is different from
galaxy-galaxy correlation not only because two different sets of objects
are being considered but also because QSOs, as opposed to galaxies,
are rare; a typical field contains of the order of $10^4$ galaxies down
to $m_R=20$,
but only a handful of bright QSOs. Therefore the various refinements
on the standard estimator discussed in the literature are not applicable
in the present case.

The estimator that we use to compute the QSO-galaxy cross-correlation
function is defined by,
\begin{equation}
\omega(\theta)={{D_Q D_G}\over{\langle R^\prime_Q D_G\rangle}}-1,
\label{eqest}
\end{equation}
where $D_Q D_G$ is the actual number of QSO-galaxy pairs of a given
separation, and
$R^\prime_Q D_G$ is the number of random QSO-real galaxy pairs. Random
QSO positions are obtained for every real QSO by randomly `scattering' 
it in the {\it azimuthal} direction with respect to the field centre, 
while keeping its radial distance the same. This estimator deals
successfully with the problem of low QSO numbers, our circular field 
boundaries, and 
radial sensitivity gradients on Schmidt plates. Had we used random 
QSOs that were scattered in both radial and azimuthal directions 
(or, equivalently in $x$ and $y$), the presence of plate edges together 
with the small number of QSOs would have resulted in spurious 
correlations. Since azimuthally scattered random QSOs sample the same 
radius-dependent galaxy density on the plates as the corresponding real 
QSOs, radial plate gradients cancel out. To remind ourselves that our 
estimator is not the standard one, we use $R^\prime_Q$, instead of $R_Q$,
in eq.~\ref{eqest}. 
The angular brackets in the denominator of eq.~\ref{eqest} mean that 
we take the average of 100 random realizations for every real QSO. 

Note that for QSOs located close to plate centres, eq.~\ref{eqest} can 
underestimate the absolute amplitude of the cross-correlation signal. 
Suppose the plate centre happens to have a real excess or deficit of 
galaxies. All the randomly generated QSOs close to field centre will 
also be sampling the same real excess or deficit of galaxies, and thus 
the cross-correlation signal will not be detected. However this is a
small effect because only a small percentage of the QSOs are close
to the plate centre.

The cross-correlation as a function of angular scale, $\theta$, is
first calculated for every field separately. The final $\omega(\theta)$ 
is the average of individual field contributions, weighted by the number
of QSOs they contain. We do not attempt to estimate correlations on
scales larger than a single plate, i.e. $\theta\simgt 4\degree$.

\begin{figure}
\vbox{
\centerline{
\psfig{figure=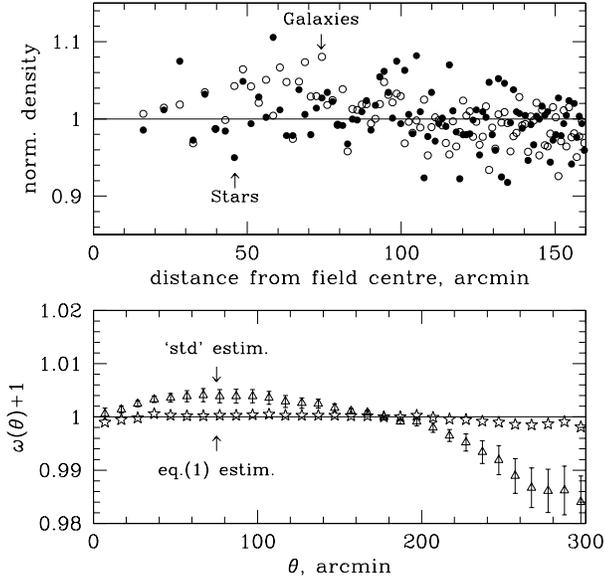,width=3.25in,angle=0}
}
\caption[]{Top panel: normalized radial number density of 
$18.5\leq m_R\leq 20.0$ APM galaxies around plate centres, summed over
11 equatorial fields (empty circles), and $17.0\leq m_R\leq 18.0$ APM 
stars (filled circles). Both types of objects show radial density
gradients. Bottom panel: cross-correlation  between stars and galaxies
using eq.~\ref{eqest} (star symbols) and `standard' (triangles)
estimators. The 1-$\sigma$ errorbars, estimated as the standard
deviation of the mean of 11 LBQS fields, are plotted in the latter case.
Note the horizontal scale in the bottom panel is twice as long as in 
the top panel, since separations as large as the plate diameter can be 
considered. 
\label{SGtest}
}
}
\end{figure}

To test our estimator we calculate the cross-correlation signal between 
galaxies (extended objects) and Galactic stars (star-like objects), both 
taken from the red APM plates. In this 
particular example galaxies are confined to a magnitude range between 
18.5 and 20.0, while stars have magnitudes between 17.0 and 18.0. 
The surface density of these objects as a function of the distance 
from plate centre, averaged over 11 plates, is shown in the top panel 
of Figure~\ref{SGtest}. The empty circles are galaxies, filled circles 
are stars. Both types of objects show radial gradients. The star-galaxy
cross-correlation estimated using eq.~\ref{eqest} is represented as star 
symbols in the lower panel of the same Figure. The cross-correlation 
estimated using the standard method, i.e. by scattering random QSOs in 
the $x$ and $y$ directions, is shown as square symbols. The estimator
defined by
eq.~\ref{eqest} has successfully compensated for galaxy and 
star radial gradients; its application shows that there is no correlation 
between stars and galaxies, as expected. On the other hand, the standard 
estimator's results are dominated by spurious effects. We therefore use 
the estimator defined by eq.~\ref{eqest} in the rest of this paper.

\subsection{Results of cross-correlation}
\label{results}

Now we estimate QSO-galaxy correlations using one subset of QSOs,
and different galaxy subsamples defined by their apparent magnitudes. 
As a check, for every QSO-galaxy correlation we calculate QSO-star 
correlation, with stars in the same magnitude range as galaxies. 

\begin{figure}
\vbox{
\centerline{
\psfig{figure=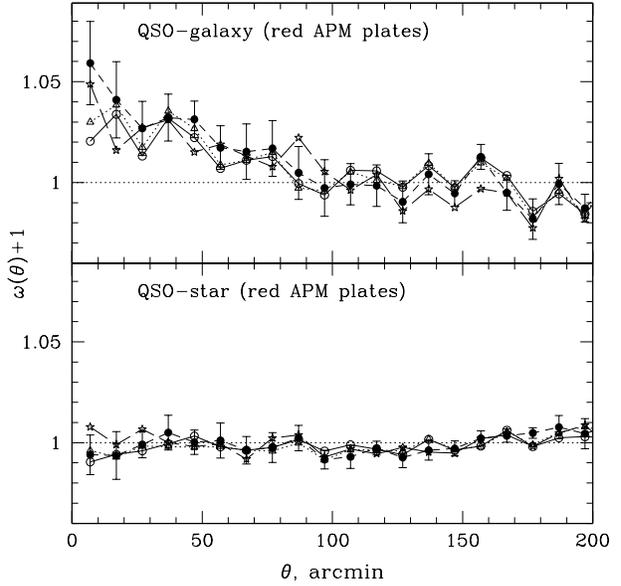,width=3.25in,angle=0}
}
\caption[]{
QSO-galaxy (top) and QSO-star (bottom) correlations using the same QSO
subsample ($z\geq 1$, $m\leq 18.0$), and four galaxy subsamples taken
from the {\it red} APM plates, with magnitude ranges: 
$16.00\rightarrow 19.50$ (empty circles);
$17.32\rightarrow 19.68$ (triangles);
$18.50\rightarrow 20.00$ (solid circles);
$19.54\rightarrow 20.46$ (stars).
Error-bars are 1$\sigma$ standard deviation of the mean of 11 LBQS
fields, and are shown only for the $18.50\leq m_R\leq 20.00$ case.
\label{tpcorr7C_rr}
}
}
\end{figure}

Figure~\ref{tpcorr7C_rr} shows QSO-galaxy
(top panel) and QSO-star (bottom panel) cross-correlation for
QSOs with $z\geq 1$ and $m\leq 18$, and four galaxy subsamples,
all taken from the red plates, with magnitude ranges:\\
$16.00\rightarrow 19.50$ (empty circles);\\
$17.32\rightarrow 19.68$ (triangles);\\
$18.50\rightarrow 20.00$ (solid circles);\\
$19.54\rightarrow 20.46$ (stars).\\
The bin widths were chosen such that the number of galaxies per bin is 
approximately the same in all bins. The QSO-galaxy cross-correlation 
signal, $\omega_{QG}(\theta)$, is detected clearly, while Galactic stars 
do not correlate with QSOs, as expected. The signal persists to 
separations of $\sim 75\arcmin$; beyond that the slope of 
$\omega_{QG}(\theta)$ becomes 
nearly zero. At $\theta \simgt 150\arcmin$ the signal is mostly negative 
because of the integral constraint imposed on the correlation function. 
Though all four galaxy magnitude bins show comparable signal, the 
strongest correlations are with $18.5\leq m_R\leq 20.0$,
i.e. the ``faintest-but-one'' bin.
Perhaps this is not surprising: if the signal is due to weak lensing
then fainter galaxy samples, i.e. those at higher redshifts, are 
expected to be better lenses for QSOs at $z\geq 1$. On the other hand, 
the faintest magnitude bin is probably contaminated by stars, which 
would tend to dilute the observed $\omega_{QG}$.

\begin{figure}
\vbox{
\centerline{
\psfig{figure=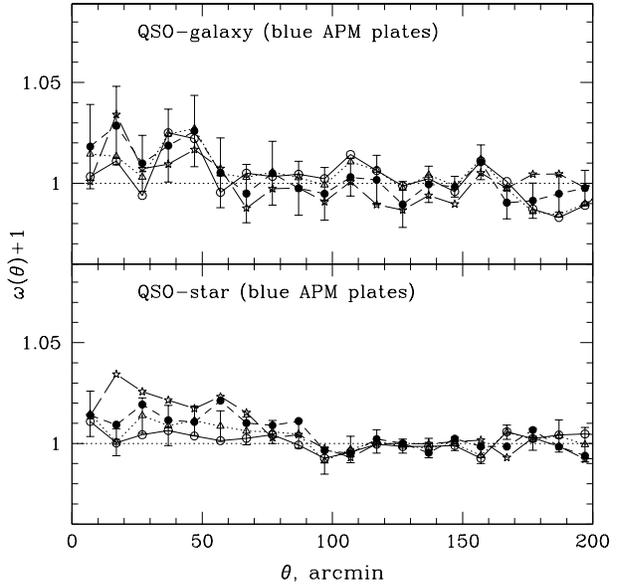,width=3.25in,angle=0}
}
\caption[]{
QSO-galaxy (top) and QSO-star (bottom) correlations using the same QSO
subsample ($z\geq 1$, $m\leq 18.0$), and four galaxy subsamples taken
from the {\it blue} APM plates, with magnitude ranges: 
$17.50\rightarrow 21.00$ (empty circles);
$18.82\rightarrow 21.18$ (triangles);
$20.00\rightarrow 21.50$ (solid circles);
$21.04\rightarrow 21.96$ (stars).
Error-bars are 1$\sigma$ standard deviation of the mean of 11 LBQS
fields, and are shown only for the $20.00\leq m_R\leq 21.50$ case.
\label{tpcorr7C_bb}
}
}
\end{figure}

Figure~\ref{tpcorr7C_bb} is the same as Figure~\ref{tpcorr7C_rr}, only 
the galaxies here were taken from the blue plates. 
The magnitude ranges are:\\
$17.50\rightarrow 21.00$ (empty circles);\\
$18.82\rightarrow 21.18$ (triangles);\\
$20.00\rightarrow 21.50$ (solid circles);\\
$21.04\rightarrow 21.96$ (stars).\\
The apparent association between `stars' on blue APM plates and QSOs
is due to a population of compact red galaxies, which appear extended 
on red plates, but are stellar on blue plates. This population has an 
average colour of $B-R=2.1$, much redder than typical galaxies in the 
18.5 to 20.0 magnitude range detected on the red plates. The existence 
of this population of galaxies is known, and preliminary spectroscopic 
work indicates that these are ellipticals at a typical redshift of 0.3. 
(Hewett et al. 1997). Recently, a faint lensed arclet of a high-redshift 
source has been observed centred on one of these galaxies (Hewett et al.
1997). This population of compact red ellipticals at $z\sim 0.3$ is also 
an important contributor to the weak lensing induced correlation between 
QSOs and galaxies. Because these galaxies appear stellar on blue APM 
plates, a positive correlation between QSOs and `stars' is detected on 
these plates (Fig.~\ref{tpcorr7C_bb}, bottom panel). Also, because this 
population is not detected as `galaxies' on blue plates, the 
corresponding QSO-galaxy correlation (Fig.~\ref{tpcorr7C_bb}, top panel) 
is weaker compared to that on the red plates. Because of these factors
affecting the blue APM plates, we will only use red plates in the rest
of the analysis.

Before we proceed, it is important to note that the amplitude of the 
signal that we measure here is a lower limit to the true signal because 
of three effects. First, as explained above (Section~\ref{estimator}) 
our cross-correlation estimator can bias the amplitude of the signal low. 
Second, because QSO candidates for the LBQS were selected using prism 
plates, candidates in crowded areas, e.g. high galaxy density, are more 
likely to be rejected as their spectra have a higher chance of being 
`corrupted' by a superimposed galaxy image. Third, APM object classifier 
and estimated object magnitudes are not perfect, and hence any true signal
which is due to galaxies in any given magnitude range will be diluted by 
stars and by objects of other apparent magnitudes.

\subsection{Integrated correlations for $\theta<75\arcmin$}
\label{75arcmin}

\begin{figure}
\vbox{
\centerline{
\psfig{figure=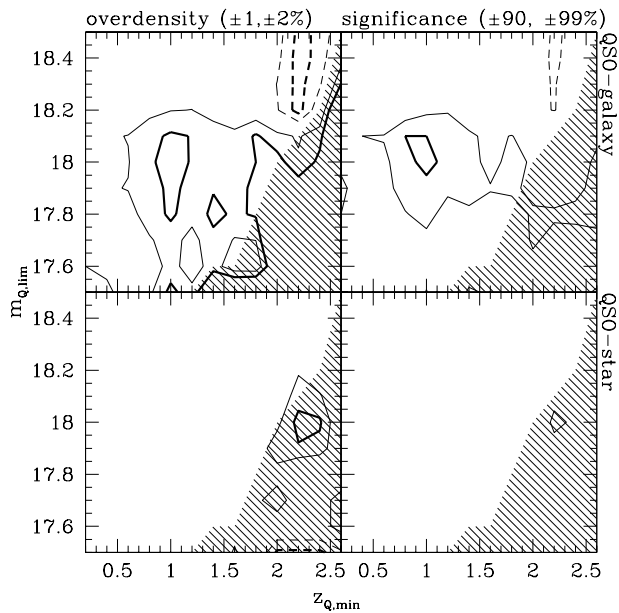,width=3.25in,angle=0}
}
\caption[]{
Contour plots of cumulative $\omega_{QG}$ and $\omega_{QS}$ for
$\theta\leq 75\arcmin$ (left), and the corresponding statistical
significance (right). The correlation contours are drawn at 
0.98, 0.99 (thick and think dashed lines), and 1.02, 1.01
(thick and thin solid lines) levels. The significance contours are at 
90, 99\% (thin and thick lines). Note that APM stars show no 
correlations with QSOs. 
\label{tpcorr7J}
}
}
\end{figure}

In the previous Section we used one subsample of QSOs. To eliminate any 
bias related to that particular choice of QSO properties, we extend the
QSO-galaxy cross-correlation
analysis to a range of QSO subsamples each having a minimum redshift
and limiting apparent magnitude, $z_{Q,min}$ and $m_{Q,lim}$. For each
subsample we calculate the integrated correlation amplitude within 
$\theta=75\arcmin$, i.e. we characterize each ($z_{Q,min}$, $m_{Q,lim}$) 
subsample by one number. This integrated correlation amplitude is 
plotted as a contour plot in Figure~\ref{tpcorr7J}.

The top two panels of Figure~\ref{tpcorr7J} show the contours of constant 
QSO-galaxy correlation amplitude (left panel) and statistical significance 
(right panel); while the bottom two panels are the same, but for QSO-star 
correlations. The correlation contours are drawn at 0.98, 0.99 (thick and 
thin dashed lines), and 1.02, 1.01 (thick and thin solid lines) levels. 
The significance contours are at 90 and 99\% (thin and thick lines). The 
significance level is the fraction of synthetic QSO subsamples, out a 
total of 100, that show correlations weaker that the corresponding real 
QSO subsample. This estimate makes no {\it a priori} assumptions about
the distribution of errors.
The right corner of each panel is shaded where the number of QSOs
per subsample is less than 10, so even if the correlations appear 
significant, they should be regarded with caution because not enough 
QSOs are being averaged over.

It is apparent from Figure~\ref{tpcorr7J} that QSO-galaxy correlations
are statistically significant for some QSO subsamples, whereas no
significant correlations, above 90\% c.l., are detected between QSOs 
and Galactic stars. Because of the null result for QSO-star and
star-galaxy (Figure~\ref{SGtest}) correlations, we conclude that 
QSO-galaxy signal is real, and not an artifact of plate sensitivity
gradients, etc. 

What are the possible causes of these correlations? The correlations
cannot be due to physical QSO-galaxy 
associations, because of the QSO/galaxy redshift mismatch. Another 
possibility is that patchy Galactic obscuration `creates' galaxy and
QSO over-densities, which are then necessarily correlated. This hypothesis
can be ruled out because the observed correlations show a strong variation
with $z_{Q,min}$ both in amplitude and significance, whereas dust would 
not be able to differentiate between QSOs at different redshifts.
Furthermore, the fluctuations in the distribution of Galactic dust are
not extreme enough to produce the observed variations in galaxy and QSO
number densities. The rms fluctuations in obscuration in these high 
Galactic latitude ($\vert b\vert\simgt 45\degree$) LBQS fields is 
typically 0.1 magnitudes in $B$ (Burstein \& Heiles 1978, 
Schlegel et al. 1997) which would produce rms fluctuations in projected
galaxy density of a factor of 1.07, while the observed rms value is
4 on $\sim10\arcmin$ scales, and 1.4 on degree scales. Thus, the observed 
galaxy density fluctuations in these APM fields are primarily due to 
galaxy clustering, and not patchy Galactic dust obscuration. 
Therefore we conclude that the correlations are not due to dust.

The only remaining explanation is the weak
gravitational lensing of the background QSOs by the matter associated
with the APM galaxies. This interpretation is supported by the 
trends seen in the top panels of Figure~\ref{tpcorr7J}. The correlations
are strongest with QSOs at $z\simgt 1$, consistent with 
theoretical expectations (Dolag \& Bartelmann 1997, Sanz et al. 1997).
The apparent magnitude-dependent behavior of the correlations is also
consistent with the lensing interpretation; $\omega_{QG}$ is strongest
for $m_Q\simlt 18$.
At fainter magnitudes the slope of the QSO luminosity 
function becomes shallower and so amplification bias, and hence 
correlation strength, become less pronounced. At magnitudes much brighter 
than $\sim 18$ the significance of the QSO-galaxy correlations is low 
due to small QSO numbers. 

\begin{figure}
\vbox{
\centerline{
\psfig{figure=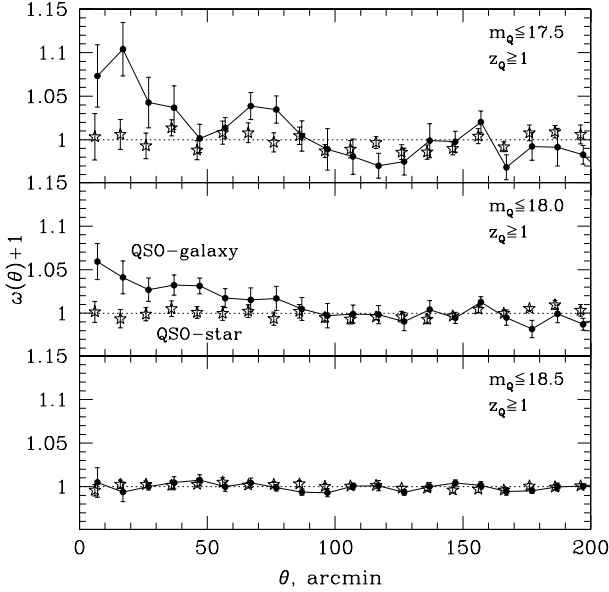,width=3.25in,angle=0}
}
\caption[]{
QSO-galaxy (solid dots and lines) and QSO-star (star symbols) 
cross-correlation function. The galaxies are taken from the red APM
plates, and have magnitudes between 18.5 and 20.0. Results for three
QSO subsamples are presented, as indicated in the upper right corner
of each panel. Note that the amplitude of $\omega_{QG}$ increases 
rapidly for brighter QSOs.
\label{tpcorr7C}
}
}
\end{figure}

As an example of the actual $\omega_{QG}$ as a function of separation,
Figure~\ref{tpcorr7C} shows the correlation between $z\geq 1$ QSOs and 
galaxies, for three different $m_{Q,lim}$: 17.5, 18.0 and 18.5. 
The star symbols are QSO-star, and solid dots are QSO-galaxy correlations.
The error-bars are 1$\sigma$ standard deviation of the mean of eleven
LBQS fields. The results shown in the middle panel, $m_Q<18$, 
are the same as solid circles in Figure~\ref{tpcorr7C_rr}. 

\section{Analysis and comparison with predictions}\label{analysis}

Having measured the weak lensing induced two-point correlation function 
between QSOs and galaxies, we now compare it to theoretical 
expectations, using two approaches (Sections~\ref{phenom} and 
\ref{literature}) that differ mostly in the method used to estimate 
the clumpiness of the mass responsible for lensing. The results of these 
two approaches are compared to each other and to the observations.
As observations, we chose the correlations presented in the middle
panel of Figure~\ref{tpcorr7C}, i.e. for a QSO subsample with 
$z_{Q}\geq 1$ and $m_{Q}\leq 18.0$. We plot these as filled circles in
Figure~\ref{tpcf}, for separations where the amplitude of $\omega_{QG}$
remains consistently positive.

\subsection{Phenomenological Predictions}\label{phenom}

The main assumption here is that the mass distribution responsible
for lensing is traced by the observed galaxies in the nearby Universe.
The total `thickness' of the lensing material can be estimated from the
galaxy redshift distribution and an assumed $\Omega_0$, while the 
distribution of fluctuations on any given scale are obtained from the
counts-in-cells distribution of the APM galaxies.

First, we estimate the amount of matter associated with the observed
APM galaxies, i.e. we calculate the optical depth for sources located
at the redshift of the QSOs,
\begin{equation}
\tau_{APM}=\rho_{crit} \Omega_0
\int_0^{z_{max}} {{(c dt/dz) (1+z)^3}\over{\Sigma_{crit}(z,z_s)}}dz,
\label{tau}
\end{equation}
Here, $cdt$ is the thickness of the lensing slice at redshift $z$, 
$\rho_{crit} \Omega_0 (1+z)^3$ is its mass density, and
$\Sigma_{crit}(z,z_s)$ is the critical lensing surface mass density 
at $z$ for a source at $z_s$. The latter is assumed to be
1.5, the average redshift of our QSO subsample.  We assume an open
Universe with $\Omega_0=0.3$. The upper limit of integration, $z_{max}$
is taken to be the redshift at which the APM galaxy redshift distribution 
(Figure~\ref{z}) drops to half of its peak value; $z_{max}$ is about 
0.33. In other words, we assume that the optical depth of the mass traced 
by the APM galaxies is the total optical depth up to a redshift of 0.33.
Since the particular choice of $z_{max}$ is somewhat arbitrary, 
we also quote results for $z_{max}$=0.28 and 0.39, corresponding to 
redshifts where the APM galaxy redshift distribution drops to 3/4 and 
1/4 of its peak value, respectively. The corresponding range of 
$\tau_{APM}$ values is from 0.012 to 0.021, while for $z_{max}$=0.33, it
is 0.016.

Next we assume that on scales larger than a few arcminutes the APM 
galaxies trace the total matter distribution (up to $z_{max}$) with a 
biasing factor $b$. The projected galaxy clustering is described by the 
scale-dependent counts-in-cells distribution of the APM galaxies,
$p(\sigma|\,\theta)\,d\sigma$, where $\sigma$ is the galaxy number 
density normalized by the average density. A patch of sky with density 
$\sigma$ produces an amplification 
$A(\sigma)\approx 1+2\tau_{APM}(\sigma-1)/b$, in the weak lensing limit. 
The corresponding over-density of QSOs background to this patch is given 
by, 
\begin{equation}
q(\sigma)=A^{2.5\alpha-1}
\approx 1+(2\,\tau_{APM}[\sigma-1]/b)\,(2.5\alpha-1),
\label{q}
\end{equation}
where $\alpha$ is the slope of 
the QSO number counts; $\langle\alpha\rangle=1.1$ for our QSO subsample.

\begin{figure}
\vbox{
\centerline{
\psfig{figure=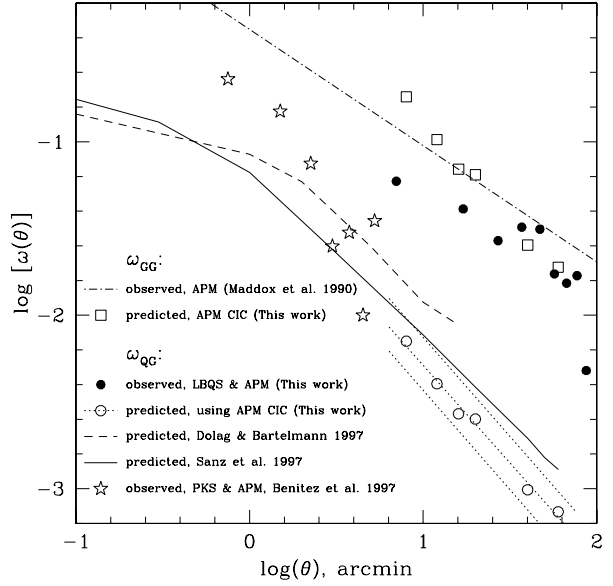,width=3.25in,angle=0}
}
\caption[]{
Observed and predicted QSO-galaxy correlations; note the log-log scale.
The different lines and points are described in the Figure. The solid
dots represent the $z\geq 1$ and $m\leq 18.0$ QSO subsample. See 
Sections~\ref{phenom} and \ref{literature} for details.
\label{tpcf}
}
}
\end{figure}

The cross-correlation between QSOs and galaxies is then estimated as
\begin{equation}
\omega_{QG}(<\theta)+1\approx
\int_0^\infty p(\sigma|\,\theta)\,q(\sigma)\,\sigma\,d\sigma.
\label{QGpredicted}
\end{equation}

The galaxy auto-correlation function is given by an equation similar 
to eq.~\ref{QGpredicted},
\begin{equation}
\omega_{GG}(<\theta)+1\approx
\int_0^\infty p(\sigma|\,\theta)\,\sigma^2\,d\sigma.
\label{GGpredicted}
\end{equation}
The ratio of $\omega_{QG}$ to $\omega_{GG}$ can be derived analytically, 
in the weak lensing regime. 
Making use of the fact that $p(\sigma|\,\theta)$ is a normalized
probability distribution with an average value of $\sigma$ equal to 1,
the ratio of
eq.~\ref{QGpredicted} and eq.~\ref{GGpredicted} gives, 
\begin{equation}
\omega_{QG}(\theta)\approx
(2\,\tau_{APM}/b)\,(2.5\alpha-1)\,\omega_{GG}(\theta),
\label{wratio}
\end{equation} 
This equation applies if the same galaxy counts-in-cells distribution
is used to calculate $\omega_{QG}(\theta)$ and $\omega_{GG}(\theta)$,
and gives the QSO-galaxy correlation amplitude due to lensing by these 
galaxies only.

For $\tau_{APM}=0.016$ (i.e. $z_{max}=0.33$), $b=1$, and 
$\langle\alpha\rangle$=1.1, the ratio 
$\omega_{QG}/\omega_{GG}\approx 0.056$. Galaxy auto-correlation
function from the APM galaxies was calculated by Maddox et al. (1990),
and is plotted as dot-dashed line in Figure~\ref{tpcf}. The amplitude 
of $\omega_{GG}$ is appropriate for
the galaxy magnitude range used here. With this
$\omega_{GG}$, QSO-galaxy cross-correlation function should lie 1.25 
(in the log) below $\omega_{GG}$. Since the observed $\omega_{QG}$ is
roughly of the same amplitude as the galaxy-galaxy correlations, the
predictions based on eq.~\ref{wratio} are a factor of 15 below 
observations. 

As a check, eqs.~\ref{QGpredicted} and \ref{GGpredicted} can be also 
applied directly to the counts-in-cells data to calculate the correlation 
functions at a range of separations, assuming the optical depth of the APM 
galaxies. The results are shown in Figure~\ref{tpcf}, as empty squares 
($\omega_{GG}$) and empty circles ($\omega_{QG}$). The vertical offset
of the two is about a factor of 15, in agreement with eq.~\ref{wratio}.
Because vertical normalization of $\omega_{GG}$ scales with the magnitude 
of the galaxies, and the data we use was not properly magnitude 
calibrated, we have adjusted the vertical normalization of these two 
estimated functions such that our derived $\omega_{GG}$ matches that of 
Maddox et al. (1990).

The empty circles assume $z_{max}$=0.33; the dotted line through them
is a rough fit. The upper and lower dotted lines are for $z_{max}$=0.39 
and 0.28 respectively, and bracket the range of our predicted values.
The model predictions cover an angular scale range of $4\arcmin$ to 
$1\degree$; on scales smaller than $4\arcmin$ the number of galaxies per 
cell is small, and so the resultant counts-in-cells distribution is 
dominated by Poisson noise. Scales much larger 
than $1\degree$ are comparable in size to the Schmidt plates, so the 
derived counts-in-cells will be substantially narrower 
then the true distribution, because we normalize the galaxy number
density to the plate's average, on each plate separately, thus 
ignoring any power on larger scales. For the same reason, the 
slope of our $\omega_{GG}(\theta)$ is somewhat steeper than the commonly 
derived value of $\gamma\approx 0.7$ (Maddox et al. 1990).

As mentioned above,
our phenomenological model predictions fall short of the observations 
by about a factor of 10--20. Is our model too simplistic?

\subsection{Theoretical Predictions from the Literature}
\label{literature}

Dolag \& Bartelmann (1997) and Sanz et al. (1997) used an alternative 
way to derive mass fluctuations in the Universe. Starting with a
particular type of cosmology and an initial matter power spectrum, 
they used the results of a semi-analytic
prescription for the non-linear evolution of the power spectrum
(Peacock \& Dodds 1996) to derive the spectrum of
mass distribution as a function of redshift and scale. Assuming weak 
lensing regime, slope of QSO counts $\alpha$, and a biasing parameter, 
the mass distribution spectrum was  used to predict the observed
QSO-galaxy correlations. These calculations were primarily motivated
by an observed correlation between PKS radio-selected QSOs and APM 
galaxies, as derived by Ben\'{\i}tez \& Mart\'{\i}nez-Gonz\'alez (1997), 
and shown as star symbols in Figure~\ref{tpcf}. Within their range of 
applicability, i.e. $\theta> 1-2\arcmin$, both analytical models 
reproduce the PKS-APM results well. Correlations on scales larger than 
$15\arcmin$ could not have been detected by Ben\'{\i}tez \& 
Mart\'{\i}nez-Gonz\'alez because they restricted their analysis to
patches of $15\arcmin$ radius around each QSO.

To compare these two model predictions to our observed 
$\omega_{QG}(\theta)$, we rescaled their results to $\alpha=1.1$.  
Dolag \& Bartelmann results were additionally rescaled from $h=0.7$ to 
$h=0.5$ using their Figure 4. The predictions are plotted as dashed and 
solid lines in Figure~\ref{tpcf}; both are for $\Omega=0.3$, $\Lambda=0$.
Similar to the results of the phenomenological model, the analytical 
models underestimate the amplitude of observed $\omega_{QG}$ by a factor 
of 10 on large angular scales.

Since phenomenological and analytical models used different routes to 
arrive at $\omega_{QG}$, it is encouraging that they agree reasonably 
well with each other. A factor of 1.5--2 discrepancy (solid and dashed
lines versus dotted lines) arises probably because our phenomenological 
model ignored lensing by structures at $z\simgt 0.35$, which are not 
sampled by galaxies in the APM Catalogue, but are still efficient lenses 
for the QSOs. Turning the argument around, the small discrepancy means 
that nearby galaxies, $z\simlt 0.35$, dominate the weak lensing of 
$z\geq 1$ QSOs.

\section{Discussion}\label{discussion}

In the last Section we have seen that the amplitude of observed 
QSO-galaxy correlation is under-predicted by theoretical models.
The discrepancy is a factor of 3--4 at $\theta\simlt 10\arcmin$, and
increases to a factor of 10 on degree scales.

Can a reasonable change of parameters account for such a large 
discrepancy? Assuming that the effects of weak lensing are dealt with 
correctly, and the signal is real and is due to lensing, we are left 
with two possibilities (see eq.~\ref{wratio}); either the slope of the 
QSO number counts at $z\geq 1$ is very steep, or mass fluctuations on 
relevant scales, i.e. between a few and 15 Mpc, are more extreme than 
fluctuations in galaxy number density.
The amplitude of $\omega_{QG}$ is quite sensitive to the slope of
the QSO number counts; however to reproduce the observations, $\alpha$ 
would have to be $\sim$ 8, while the observed slope values for LBQS QSO 
in this redshift range lie between 0.8 and 1.5, comparable to the
QSO luminosity function slopes, $\sim 1.6$, derived from the LBQS 
(Hewett et al. 1993) in this redshift range.

Alternatively, we could require that the galaxies are anti-biased with 
respect to the mass by a factor of $\sim$ 10, ($b\sim$~0.1) implying a 
$\sigma_8$ value of around 10, which is hard to reconcile with any 
gravitational clustering scenario, and observations of the large scale 
structure. The value of $\Omega^{0.6}\sigma_8$ is well constrained by 
the abundance of galaxy clusters; White, Efstathiou \& Frenk (1993) 
estimate $\Omega^{0.6}\sigma_8$ to be $\sim 0.6$. Based on cluster
peculiar velocities, Watkins (1997) derives $\Omega^{0.6}\sigma_8$ to 
be 0.44. Thus an $\sigma_8\sim 10$ would require an $\Omega$ so low, 
it would just agree with $\Omega_{baryon}$ from primordial 
nucleosynthesis constraints (Hogan 1996), but would strongly contradict 
dynamical measurements on cluster scales ($\Omega$=0.24, Carlberg et al.
1996), and power spectrum shape parameter constraints 
($\Gamma$=$\Omega\,h$=0.2--0.3, Maddox et al. 1996). Furthermore, such a 
high $\sigma_8$ value is ruled out by the recent direct estimation by 
Fan, Bahcall \& Cen (1997), who compute $\sigma_8=0.83\pm 0.15$ from
the redshift evolution of galaxy clusters. 

A combination of factors, for example, $b$=0.3 and $\alpha$=2.3 shares 
the burden between the two observables, but is still quite unpalatable.
In fact, it is difficult to see how any reasonable change of parameters 
or assumptions can bring these up to the observed amplitude of 
$\omega_{QG}$ on $\sim$ degree scales. 

If the explanation does not lie with either the lensers or the lensees,
then maybe the lensing process has to be looked at more carefully. It
is conceivable, for example, that some hitherto unexplored non-linearities 
in the description of light propagation through a clumpy universe 
could be responsible.

At present, the factor of 10 discrepancy between observations and 
predictions remains unaccounted for.

\section {Implications for Observational Cosmology}\label{implications}

The results of the correlations described above have interesting 
implications for the observed distribution and properties of certain 
types of high-redshift sources. Here we explore a few of these.

\begin{figure}
\vbox{
\centerline{
\psfig{figure=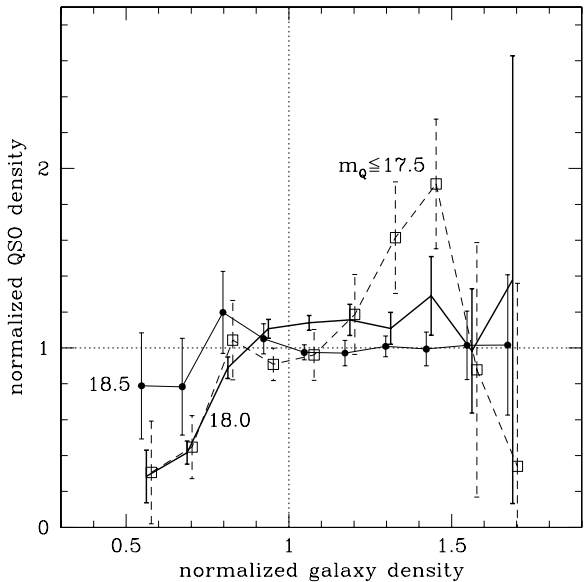,width=3.25in,angle=0}
}
\caption[]{
Normalized QSO density for a given normalized galaxy density, in circular
patches of radius $12\arcmin$. See Section~\ref{implications} for details. 
Galaxies are $18.5\leq m_R\leq 20.0$; QSOs are at $z\geq 1$. Bright QSOs 
are more likely to be found in the directions over-dense in nearby 
galaxies, and avoid direction of lower than average galaxy densities.
\label{fdensity6}
}
}
\end{figure}

The results imply that the projected 
density of foreground structure affects the density of high 
redshift objects. To quantify this statement in the case of QSOs
our results can be represented 
as a plot of QSO number density as a function of foreground 
galaxy density on a given angular scale. To do that, we randomly lay down
a large number of circular patches over LBQS fields. In each patch we 
calculate galaxy number density. For all patches of a given galaxy 
density we then calculate the average QSO density, and repeat the 
process for several values of galaxy density. The results are shown in 
Figure~\ref{fdensity6}, using patches of radius $12\arcmin$, and three
QSO subsamples, all with $z_Q\geq 1$. The error-bars are $1\sigma$
standard deviations of the mean of 10 different realizations of the
experiment just described. 

It is seen that the observed projected density of QSOs is a function
of the foreground galaxy density, in the sense that bright QSOs are
found preferentially in the directions of over-dense galaxy regions;
for example, a $m_Q\leq 17.5$, $z_Q\geq 1$ optically selected QSO is 
twice as likely to be found in the direction where the galaxy density 
is 1.5 times the average. However, QSOs are not the only class of objects 
affected by weak lensing on large scales. Any population of high-redshift 
objects, whose number counts slope is substantially different from 0.4 
will be affected in a similar fashion. Intrinsically bright galaxies, 
those on the exponential part of the Schechter luminosity function 
(Schechter 1976), confined to a narrow redshift range, are an example.
As indicated in Figure~\ref{fdensity6}, both under- and over-densities
of background sources can be expected. Thus, this bears direct relevance
to the Hubble Deep Field (R. E. Williams et al. 1994), which was 
intentionally chosen to lie in the direction of the sky devoid of nearby,
$z\simlt 0.3$, structure (the nearest galaxy cluster is $48\arcmin$ away, 
etc). Depending on the exact under-density of nearby galaxies in the HDF, 
and the intrinsic luminosity of high-redshift galaxies, the latter can 
be depleted by a factor of up to 5.

A further implication is that the rms dispersion in the observed
luminosities of standard candles, and observed sizes of standard rulers
at high-redshift would be somewhat larger than currently predicted from 
numerical simulations of the lensing effects (Wambsganss et al. 1997). 
A rough extrapolation from our results shows that the rms dispersion in 
amplifications would be $0.1-0.2$ magnitudes, comparable to the 
light-curve shape corrected dispersion in the luminosities of four
$z\sim 1$ supernovae of Type Ia (Garnavich et al. 1997), but smaller 
then the dispersion in sizes of compact radio sources (Kellerman 1993) 
and double-lobed radio sources (Buchalter et al. 1997). The fact that
the rms dispersion in these observed quantities is comparable to or larger 
than what is implied by our weak lensing analysis, means that the rms 
spread in lensing induced amplifications is not usefully constrained by 
the current samples of standard candles and rulers. Because the weak 
lensing induced dispersion in the observed quantities is not Gaussian 
(or even symmetric), the overall trend with redshift can also be affected
(also see Wambsganss et al. 1997).

Another aspect of extragalactic astronomy that is affected by weak
lensing is the density and redshift evolution of Ly-$\alpha$ forest
clouds. It is evident from Figure~\ref{fdensity6} that bright, high 
redshift QSOs do not sample random lines of sight, as is generally hoped 
in the studies of intervening QSO Ly-$\alpha$ absorption lines. Since the 
most efficient lenses for a wide range of source redshifts are at 
$z_l\simlt 0.6$, one would expect a corresponding increase in the detected 
density of high equivalent width
%%%$W_\lambda \simgt 0.5\AA$, 
Ly-$\alpha$ forest clouds, which are known to be loosely associated with 
galaxies (Stocke et al. 1995, Bowen et al. 1996, Lanzetta et al. 1995)
at these redshifts. An increase in d$N$/d$z$ at $z\simlt 1$ compared to a 
power law extrapolation from higher redshifts has been detected 
(Bahcall et al. 1993, Impey et al. 1996); a substantial contribution to 
this increase is probably due to weak lensing of back-lighting QSOs.

\section{Conclusions}\label{conclusions}

We have found that bright, high-redshift, optically selected LBQS QSOs 
are positively correlated with foreground APM galaxies. 
The cross-correlations are significant and are not physical in nature, 
or due to patchy Galactic obscuration. The most plausible explanation,
which is also supported by the qualitative behavior of the correlations,
is weak gravitational lensing of the background QSOs by the intervening
dark matter traced by the APM galaxies. However, the amplitude of 
correlations on degree scales, or $\sim$ 10-15 Mpc at the redshift of
the lenses, is a factor of 10 higher than predicted from models.
The discrepancy is hard to reconcile with what we currently believe to 
be true about the Universe.

The implications of the effects of weak lensing for the observed high 
redshift Universe are far reaching, therefore it is imperative to study
such correlations further using other large uniform data sets. An
analysis with radio selected high-redshift sources would be ideal, as
these sources are immune to the effects of Galactic dust obscuration.
Samples of high-redshift sources, both optical and radio, observed down 
to faint flux limits would be useful, as both positive correlations and
anti-correlations with foreground galaxies are expected, depending on 
the limiting flux of the source subsample.

Finally, it is interesting to note that the QSO-galaxy correlations 
described here are probing the physical scale where the galaxy power 
spectrum is observed to have a kink, corresponding to a primordial
feature in the true linear power spectrum, which is not reproduced by 
any variants of the CDM model (Peacock 1997).

\section{Acknowledgments}
LLRW would like thank Steve Maddox for useful discussions on the APM
Catalogue and correlation functions. LLRW acknowledges the support of 
PPARC fellowship at the Institute of Astronomy, Cambridge, UK.

\end{document}